\documentclass[3p,times]{elsarticle}
\usepackage{ecrc}
\usepackage{epstopdf}
\usepackage{amssymb}
\usepackage{graphicx}
\usepackage{latexsym}
\usepackage{epsfig}
\usepackage{amsmath}
\usepackage{bbm}
\usepackage{color}
\usepackage{hyperref}
\volume{00}
\firstpage{1}
\journalname{Nuclear Physics A}
\runauth{M. Panero et al.}
\jid{nupha}
\jnltitlelogo{Nuclear Physics A}
\usepackage{graphicx}
\usepackage{amsmath,amssymb}
\biboptions{sort&compress}

\newcommand{\SU}{\mathrm{SU}}

\newcommand{\dd}{{\rm{d}}}

\newcommand{\tr}{{\rm Tr}}

\newcommand{\qhat}{\hat{q}}

\newcommand{\gE}{g_{\mbox{\tiny{E}}}}

\newcommand{\mE}{m_{\mbox{\tiny{E}}}}
\newcommand{\mD}{m_{\mbox{\tiny{D}}}}
\newcommand{\Cfund}{\mathcal{C}_{\mbox{\tiny{f}}}}
\newcommand{\Cadj}{\mathcal{C}_{\mbox{\tiny{a}}}}

\newcommand{\eq}[1]{\begin{equation}\label{#1}}
\newcommand{\en}{\end{equation}}
\newcommand{\eqar}[1]{\begin{eqnarray}\label{#1}}
\newcommand{\enar}{\end{eqnarray}}

\begin{document}

\begin{frontmatter}

\title{Jet quenching from the lattice}

\author[IFT]{Marco~Panero}
\author[HY]{Kari~Rummukainen}
\author[UR]{Andreas~Sch\"afer}

\address[IFT]{Instituto de F\'{\i}sica T\'eorica, Universidad Aut\'onoma de Madrid \& CSIC, E-28049 Cantoblanco, Madrid, Spain}
\address[HY]{Department of Physics \& Helsinki Institute of Physics, P.O. Box 64, FI-00014 University of Helsinki, Finland}
\address[UR]{Institute for Theoretical Physics, University of Regensburg, D-93040 Regensburg, Germany}

\begin{abstract}
We present a lattice study of the momentum broadening experienced by a hard parton in the quark-gluon plasma. In particular, the contributions to this real-time phenomenon from soft modes are extracted from a set of gauge-invariant operators in a dimensionally reduced effective theory (electrostatic QCD), which can be simulated on a Euclidean lattice. At the temperatures accessible to present experiments, the soft contributions to the jet quenching parameter are found to be quite large. We compare our results to phenomenological models and to holographic computations.
\end{abstract}


\end{frontmatter}

\section{Introduction}

Jet quenching, namely the suppression of particles with large transverse momenta and of correlations between back-to-back hadrons detected after a heavy-ion collision, is an effect directly related to the energy loss and momentum broadening experienced by a hard parton moving in the deconfined medium, due to its interactions with the quark-gluon plasma (QGP) constituents~\cite{Bjorken:1982qr}.

Under the assumption that the parton is much harder than the typical momenta of thermal excitations in the QGP, the standard formalism to describe jet quenching theoretically relies on a multiple soft-scattering picture, in the eikonal approximation~\cite{Baier:1996kr, Baier:1996sk, Baier:2000mf, Kovner:2003zj, CasalderreySolana:2007zz}. The average increase in the (squared) transverse momentum component of the hard parton per unit length is constant, and defines the phenomenological jet quenching parameter $\qhat$,
\begin{equation}
\label{qhat_definition}
\qhat = \frac{\langle p^2_\perp \rangle}{L} = \int \frac{ \dd^2 p_\perp}{(2\pi)^2} p^2_\perp C(p_\perp),
\end{equation}
expressed as the second moment of the differential collision rate between the parton and the QGP constituents, $C(p_\perp)$. In turn, the latter quantity is directly related to the two-point correlation function of Wilson lines on the light cone.

What tools can be used to calculate this two-point correlator of null Wilson lines? Analytical weak-coupling expansions are a well-defined first-principles approach; however, the infrared divergences characteristic of thermal QCD pose limitations on the order to which they can be pushed~\cite{Linde:1980ts, Gross:1980br}---and the quantitative accuracy of perturbative computations truncated at the leading (LO) or next-to-leading order (NLO) is generally observable-dependent, and may be questionable at RHIC and LHC temperatures $T$, at which the QCD coupling $g$ is not very small~\cite{Laine:2005ai}. On the other hand, holographic computations based on the gauge/string correspondence are an ideal tool to investigate the strong-coupling limit of the plasma; however, they are not derived from the microscopic formulation of QCD, but rather from some models, like the $\mathcal{N}=4$ supersymmetric Yang-Mills theory~\cite{CasalderreySolana:2011us}. Finally, numerical lattice calculations (which do not rely on either strong- or weak-coupling assumptions) are based on a Euclidean formulation, hence they are generally unsuited for the whole class of phenonomena involving real-time dynamics in the QGP~\cite{Meyer:2011gj}.

\section{Soft contributions from lattice EQCD}

As pointed out in ref.~\cite{CaronHuot:2008ni} (see also ref.~\cite{Ghiglieri:2013gia}), however, it is possible to show that the contribution to $C(p_\perp)$ from \emph{soft} QGP modes (i.e., those at momentum scales up to $gT$) can be exactly evaluated in a dimensionally reduced, low-energy effective theory, namley electrostatic QCD (EQCD)~\cite{Appelquist:1981vg, Nadkarni:1982kb, Nadkarni:1988fh, Braaten:1995cm, Braaten:1995jr, Kajantie:1995dw, Kajantie:1997pd, Kajantie:1997tt}, which is nothing but Yang-Mills theory in three spatial dimensions, coupled to an adjoint scalar field. The EQCD Lagrangian is
\begin{equation}
\mathcal{L} = \frac{1}{4} F_{ij}^a F_{ij}^a + \tr \left( (D_i A_0)^2 \right) + \mE^2 \tr \left( A_0^2 \right) + \lambda_3 \left( \tr \left( A_0^2 \right) \right)^2 ;
\end{equation}
its parameters can be fixed by matching to high-temperature QCD. For example, at LO the gauge coupling, the squared mass and the quartic coupling of the scalar are related to the QCD parameters via
\begin{equation}
\gE^2 = g^2 T + \dots , \qquad
\mE^2 = \left( 1+\frac{n_f}{6}\right)g^2 T^2 + \dots , \qquad 
\lambda_3 = \frac{9- n_f}{24 \pi^2}g^4T + \dots , 
\end{equation}
where $n_f$ denotes the number of dynamical light quark flavors. This effective theory can be regularized on a lattice~\cite{Hietanen:2008tv} and studied non-perturbatively by means of Monte Carlo simulation. The parameters of our study correspond to QCD with $n_f=2$ light quarks at $T \simeq 398$~MeV and at $T \simeq 2$~GeV (roughly equal to twice and ten times the deconfinement temperature). To get sufficient accuracy at these ``low'' temperatures, we included subleading corrections in the EQCD parameter definitions.

Although this effective theory is purely spatial, the operator of interest for our computation of $\hat{q}$ must describe dynamical evolution \emph{in real time}~\cite{Panero:2013pla}. This operator can be interpreted as the dimensionally-reduced counterpart of (a gauge-invariant version of) the light-cone Wilson line correlator, and can be written as the trace of a ``decorated Wilson loop'':
\begin{equation}
\label{W_definition}
W(\ell,r) = \tr \left( L_3(x, \ell ) L_1 ( x + \ell \hat{3}, r ) L^{-1}_3(x+r\hat{1}, \ell ) L^\dagger_1(x,r)\right) ,
\end{equation}
having denoted the point at which the loop starts (and ends) as $x$, the direction of the spatial component of the light-cone Wilson lines as $\hat{3}$, and the direction of the spatial separation between the lines as $\hat{1}$, with
\begin{equation}
L_3( x, \ell ) = \prod_{n=0}^{\ell/a -1} U_3\left( x + a n \hat{3}\right) H \left( x + a (n+1) \hat{3} \right), \qquad H(x)=\exp[- a \gE^2 A_0(x)], \qquad
L_1( x, r ) = \prod_{n=0}^{r/a-1} U_1\left( x + a n \hat{1}\right) .
\end{equation}
Note that $H(x)$ represents a parallel transporter along a \emph{real-time} interval of length equal to the lattice spacing $a$, and is a \emph{Hermitian} (rather than unitary) matrix. The $W$ operator enjoys well-defined renormalization properties~\cite{D'Onofrio:2014qxa}.

\section{Numerical results}

The exponential decay of $\langle W(\ell,r) \rangle \simeq \exp \left[ - \ell V(r) \right]$ at large $\ell$ can be studied accurately using a multivelel algorithm~\cite{Luscher:2001up} and defines the quantity $V(r)$, which equals minus the transverse Fourier transform of the collision kernel $C(p_\perp)$ (up to a constant). Eq.~(\ref{qhat_definition}) implies that (the soft contribution to) the jet quenching parameter $\qhat$ is related to the \emph{curvature} of $V(r)$ near the origin. Fitting our lattice results for $V(r)$ to a functional form which includes linear, quadratic, and logarithmic-times-quadratic terms (and including the contribution from hard modes, which can be reliably computed perturbatively and is numerically subdominant) we get a final estimate for $\qhat$ around $6$~GeV$^2$/fm for $T \simeq 398$~MeV (i.e. at a temperature comparable to those realized at RHIC), with total uncertainty around $15$--$20 \%$.

\begin{figure}
\begin{center}
\includegraphics*[width=0.49\textwidth]{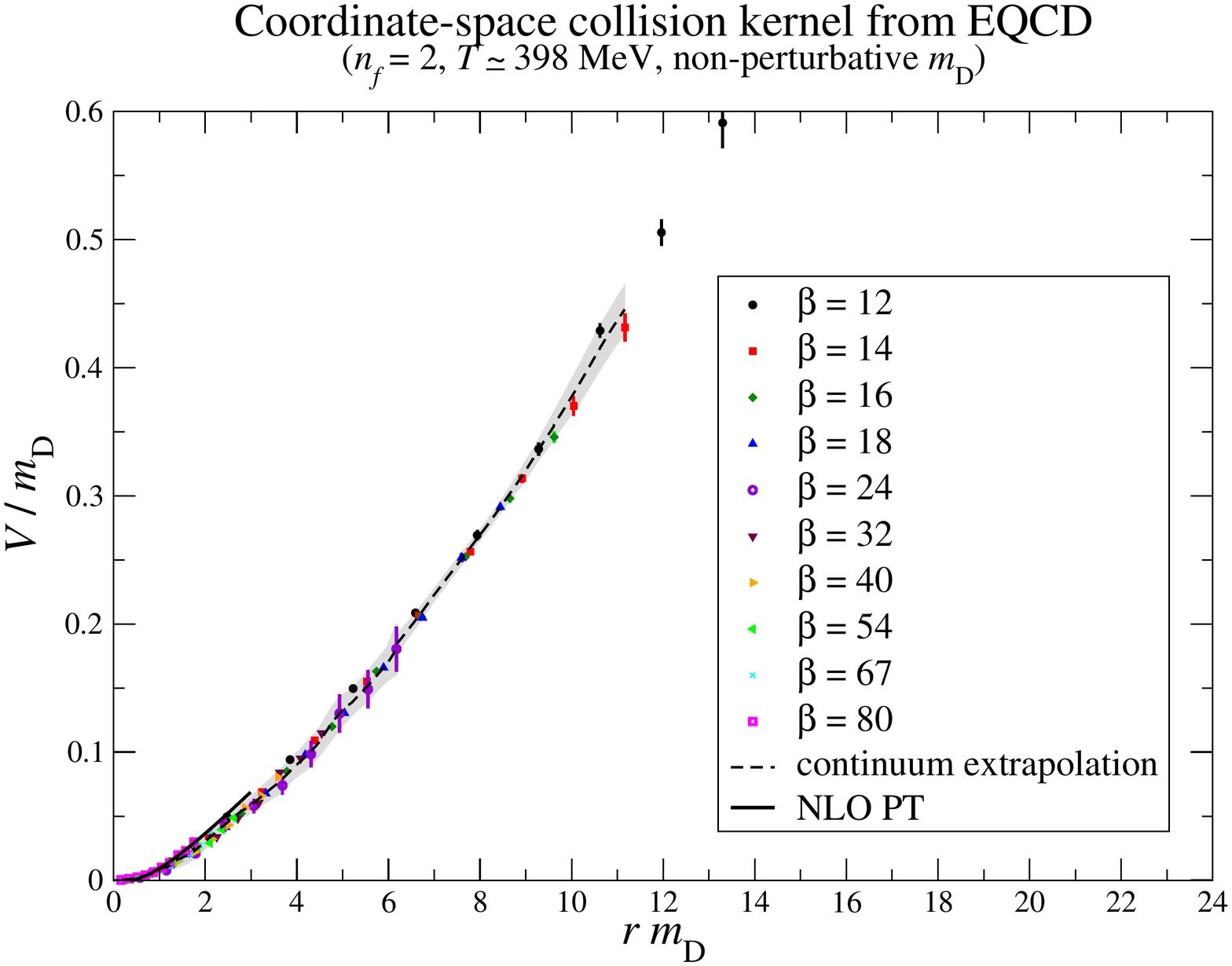} \hfill \includegraphics*[width=0.49\textwidth]{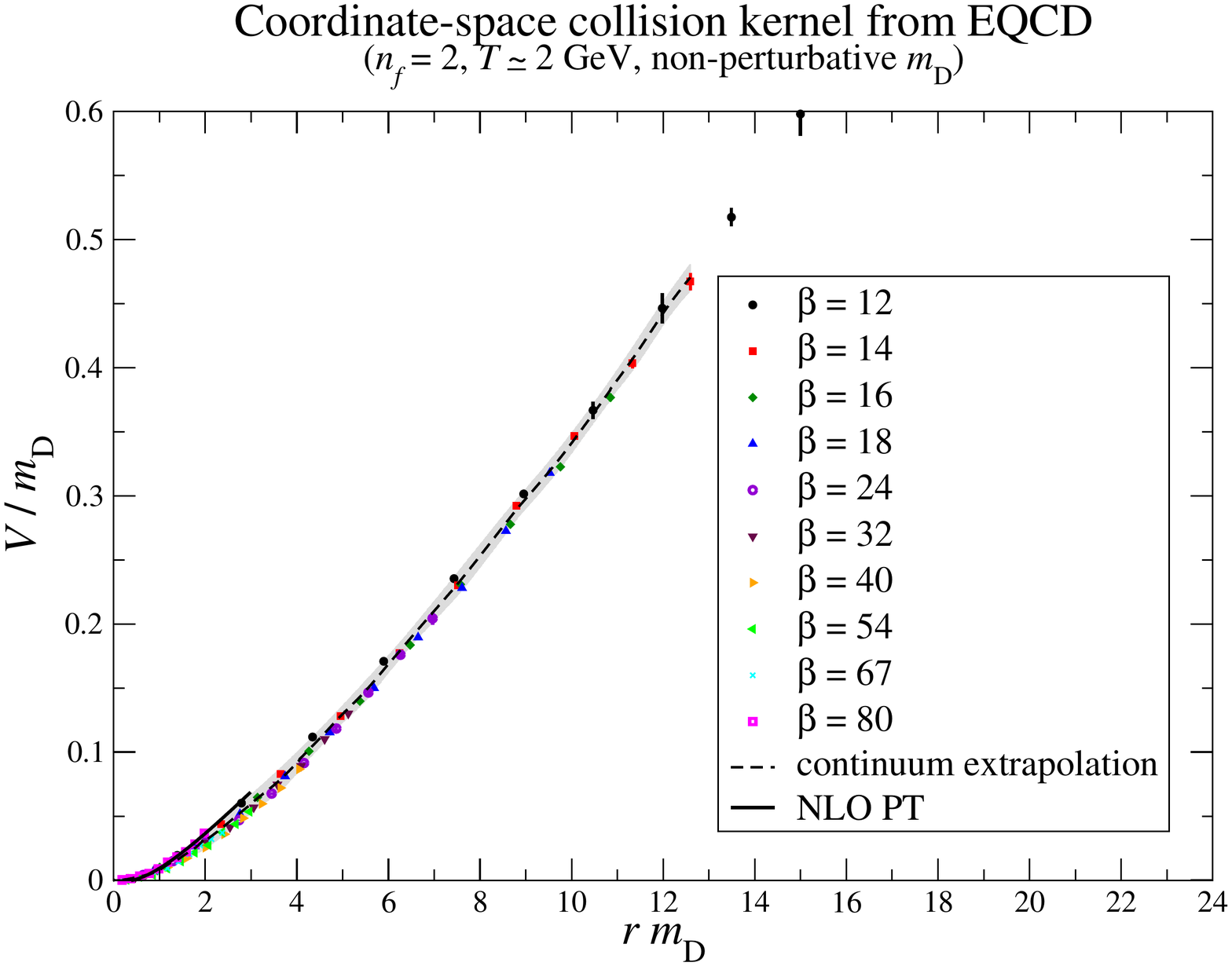}
\caption{The coordinate-space collision kernel $V(r)$ computed non-perturbatively in EQCD simulations, at $T \simeq 398$~MeV (left-hand-side panel) and at $T \simeq 2$~GeV (right-hand-side panel), in units of the non-perturbative Debye screening mass $\mD$~\cite{Laine:1999hh}. Symbols of different colors correspond to simulations at different lattice spacings $a$, with $\beta = 6/(a \gE^2)$. The dashed black line (and the gray band) show the continuum extrapolation (and the corresponding uncertainty), while the solid black curve is the perturbative prediction at NLO~\cite{CaronHuot:2008ni, Ghiglieri:2013gia}.\label{fig:V}}
\end{center}
\end{figure}

This result indicates that the non-perturbative contribution to $\qhat$ from soft modes is non-negligible, and significantly larger than expected from a na\"{\i}ve parametric analysis in perturbation theory. It is interesting to note that the mismatch between our non-perturbative results and the perturbative NLO predictions~\cite{CaronHuot:2008ni, Ghiglieri:2013gia} can be related to the existence of large non-perturbative contributions to the Debye mass $\mD$: as shown in fig.~\ref{fig:V}, plotting our results for $V(r)$ in units of the non-perturbatively estimated Debye mass~\cite{Laine:1999hh} brings our results in agreement with the curve predicted perturbatively at NLO, and makes the curves obtained at the two different temperatures compatible with each other (within uncertainties). Plugging the value of the non-perturbative Debye mass into the analytical expression for $\qhat$
\begin{equation}
g^4 T^2 \mD \Cfund \Cadj \frac{ 3 \pi^2 + 10 - 4 \ln 2 }{32 \pi^2} 
\end{equation}
(where $\Cfund=4/3$ and $\Cadj=3$ denote the eigenvalues of the quadratic Casimir operators for the fundamental and for the adjoint representation of $\SU(3)$) results, again, in a final value of $\qhat$ around $6$~GeV$^2$/fm at RHIC temperatures.

\section{Discussion and conclusions}

In this contribution, we reported on our recent lattice study of the momentum broadening experienced by a light quark in the QGP~\cite{Panero:2013pla}. Our computation is based on the idea of separating the contribution from hard thermal excitations (which can be evaluated analytically in a weak-coupling calculation) from those due to modes up to the soft scale, which we extracted non-perturbatively from Monte Carlo simulations of a dimensionally reduced, low-energy effective theory, EQCD. Related studied have also been carried out in magnetostatic QCD (which describes the physics of ``ultrasoft'', $O(g^2T/\pi)$, modes of thermal QCD)~\cite{Benzke:2012sz, Laine:2012ht}, where it was found that the contribution to $\qhat$ from the ultrasoft scale was essentially negligible. By contrast, our results indicate that, at least at experimentally accessible temperatures, non-perturbative contributions in the soft sector are non-negligible. In particular, our final result for $\qhat$ at RHIC temperatures is around $6$~GeV$^2$/fm. This value is close to estimates obtained from holographic studies~\cite{Liu:2006ug, Armesto:2006zv, Gursoy:2009kk}, and also from certain phenomenological model computations~\cite{Dainese:2004te, Eskola:2004cr}. Although more recent studies of this type tend to favor smaller values~\cite{Burke:2013yra}, one should note that a quantitative comparison is difficult, because the precise numerical value of $\qhat$ depends on details of the kinematics that is assumed. Interestingly, we also found that, by expressing our results for $V(r)$ (the collision kernel in transverse coordinate space) in units of the non-perturbatively evaluated Debye mass $\mD$ brings our results to agree with the perturbative calculation.

The approach underlying our computation allows one to bypass the intrinsic challenges of \emph{ab initio} studies of real-time phenomena on a lattice with Euclidean signature, following the seminal observation~\cite{CaronHuot:2008ni} that soft contributions to light-cone physics can be \emph{exactly} computed in the purely spatial (and bosonic) effective theory describing the thermal excitations of the QGP up to momenta $O(gT)$. A closely related observation is that the screening masses of the QGP can be related to light-cone real-time rates~\cite{Brandt:2014uda}. An explicit check of the fact that the soft contribution to $C(p_\perp)$ can be extracted ``crossing the light cone'' was carried out in classical lattice gauge theory in ref.~\cite{Laine:2013lia}.

The approach followed in the present work could be used to investigate various other real-time phenomena on the lattice. For quantities requiring a delicate control of lattice discretization effects, it might be suitable to resort to improved lattice actions, for which sophisticated error-reduction algorithms already exist~\cite{Mykkanen:2012dv}.

Other interesting extensions of this study include a more detailed investigation of the dependence of $\qhat$ on the temperature (beyond the purely dimensional expectation $\qhat \propto T^3$) and on the number of color charges $N$. The latter plays an important r\^ole in the context of holographic computations (see ref.~\cite[subsect.~2.6]{Lucini:2012gg} and references therein), hence it would be important to check if quantities related to real-time dynamics in thermal QCD also exhibit a mild dependence on $N$, as equilibrium quantities  do~\cite{Lucini:2002ku, Panero:2009tv, Datta:2009jn, Mykkanen:2012ri}.

\vskip0.1cm 
\noindent{\bf Acknowledgements} This work is supported by the Spanish MINECO (grant FPA2012-31686 and ``Centro de Excelencia Severo Ochoa'' programme grant SEV-2012-0249), by the Academy of Finland (project 1134018), by the German DFG (SFB/TR 55), and partly by the European Community (FP7 programme HadronPhysics3).

\bibliographystyle{elsarticle-num}
\bibliography{paper}

\end{document}